\documentclass[onecolumn]{article}
\usepackage{graphicx}
\usepackage{amsmath}

\input{tcilatex}

\begin{document}

\title{Optical implementability of the two--dimensional Quantum Walk}
\author{Eugenio Rold\'{a}n and J.C. Soriano \\
Departament d'\`{O}ptica, Universitat de Val\`{e}ncia, \\
Dr. Moliner 50, 46100--Burjassot, Spain}
\maketitle

\begin{abstract}
We propose an optical cavity implementation of the two--dimensional coined
quantum walk on the line. The implementation makes use of only classical
resources, and is tunable in the sense that a large number of different
unitary transformations can be implemented by tuning some parameters of the
device.
\end{abstract}

\section{Introduction}

The quantum walk (QW) is an interesting quantum process that is attracting
much attention from the algorithmic point of view \cite{Ambainis04}, but
also because of its intrinsic interest \cite{Kempe03} through its connection
with quantum cellular automata \cite{Meyer96}, and with the physics of the
systems in which it can be implemented. Two different types of QWs have been
introduced, the so--called discrete and continuous QWs. The discrete QW can
be thought of as a quantum version of the classical quantum walk \cite%
{Meyer96,Aharonov93}, whilst the continuous QW is a quantum generalization
of the Markov chain \cite{Fahri98}. In this article we shall deal only with
the discrete QW.

As stated, the discrete QW can be shortly defined as a quantum counterpart
of the random walk. In the random walk on the line, the "walker" moves to
the right or to the left depending on the output of some random process,
e.g., the toss of a coin. In the QW, the classical coin is substituted by a
quantum one, a qubit, and the coin toss is replaced by some unitary
operation acting on the qubit state, e.g, a Hadamard transformation. After
the unitary operation, the qubit state is in a superposition state and thus
there is a finite probability amplitude for the walker to move, in the same
step, to the left and to the right. This leads to the appearance of
interference phenomena in the probability distribution of the walker
localization that makes it very different from its classical counterpart.

The coined QW in one dimension has been studied extensively along the recent
years \cite%
{Nayak,Konno,Carteret,Kendon03,Lopez03,Knight04,Feldman04,Romanelli04,Romanelli05}%
, and some generalizations of the basic process have been recently proposed 
\cite%
{Wojcik04,Inui04,Buerschaper04,Ribeiro04,Romanelli04(b),Omar04,Venegas04}.
Regarding physical implementations, there are a number of proposals that
consider quantum systems, i.e., systems whose dynamics can be described only
within the framework of quantum mechanics \cite%
{Travaglione,Dur,Sanders,Zhao,Di04}. Interestingly enough, the
one--dimensional QW has been shown to be implementable by only classical
means, i.e., in setups whose description does not require quantum mechanics 
\cite{Hillery,Knight03,Knight03(b),Jeong}; and, in fact, it has been nearly
implemented in an optical cavity \cite{Bouwmeester}, as it is shown in \cite%
{Knight03,Knight03(b)}. Moreover, it has been claimed that the
one--dimensional QW is an interference phenomenon in which entanglement, a
distinctive quantum feature, does not play any role \cite{Knight03} (see
also \cite{Kendon05} for a different view).

Of course, as it is the case for the random walk, the QW can be defined in a
space of arbitrary dimensionality \cite{Mackay02}. In the multidimensional
case, in which the particle "walks" in a $d$--dimensional space, a qubit is
necessary for each spatial dimension or, in other words, a $d$--dimensional
QW requires a qu$d$it. This makes that the unitary transformations, the
analogous to the coin toss, be more complex that in the unidimensional case.
Multidimensional QWs have been studied in some detail in \cite%
{Tregenna03,Inui04(b)} but, to the best of our knowledge, no proposal for
its implementation is available to this day. In this article, we propose a
way for implementing the two--dimensional quantum walk in an optical cavity.

\section{Two--dimensional quantum walk}

Let us briefly introduce the two--dimensional QW, whose implementation is
our main goal. Consider a single particle (the walker) and a qu$d$it with
four states that plays the role of the coin. Notice that the qu$d$it can
correspond to internal states of the particle, although not necessarily. Let 
$\mathcal{H}_{P}$ be the Hilbert space of the particle positions on the
plane and 
\begin{equation}
\left\{ \left\vert x,y\right\rangle =\left\vert x\right\rangle \left\vert
y\right\rangle ,x,y\in \mathrm{Z}\right\} ,
\end{equation}%
a basis of $\mathcal{H}_{P}$; and let $\mathcal{H}_{C}$ be the
four--dimensional Hilbert space describing coin--qu$d$it, and $\left\{
\left\vert u\right\rangle ,\left\vert d\right\rangle ,\left\vert
r\right\rangle ,\left\vert l\right\rangle \right\} $ a basis of $\mathcal{H}%
_{C}$. The state of the total system belongs to the space $\mathcal{H}=%
\mathcal{H}_{C}\otimes \mathcal{H}_{P}$, and at a given instant of time, say
at iteration $n$, can be expresed as 
\begin{equation}
\left\vert \psi \right\rangle _{n}=\sum_{x,y}\left[ r_{x,y}^{\left( n\right)
}\left\vert x,y,r\right\rangle +l_{x,y}^{\left( n\right) }\left\vert
x,y,l\right\rangle +u_{x,y}^{\left( n\right) }\left\vert x,y,u\right\rangle
+d_{x,y}^{\left( n\right) }\left\vert x,y,d\right\rangle \right] ,
\end{equation}%
where the notation is self--explicative.

The dynamics of the system is governed by two physical operations: (i), the
conditional displacement, represented by the operator $\hat{D}$ acting on $%
\mathcal{H}_{P}$ 
\begin{eqnarray}
\hat{D}\left\vert x,y,r\right\rangle &=&\left\vert x+1,y,r\right\rangle ,\ \
\ \ \hat{D}\left\vert x,y,l\right\rangle =\left\vert x-1,y,l\right\rangle ,
\label{D1} \\
\hat{D}\left\vert x,y,u\right\rangle &=&\left\vert x,y+1,u\right\rangle ,\ \
\ \hat{D}\left\vert x,y,d\right\rangle =\left\vert x,y-1,d\right\rangle ,
\label{D4}
\end{eqnarray}%
i.e., the walker is displaced up, down, rigth or left when the coin is in
the state $\left\vert r\right\rangle $, $\left\vert l\right\rangle $, $%
\left\vert u\right\rangle $, or $\left\vert d\right\rangle $, respectively;
and (ii), the unitary transformation acting on the internal states of the
coin, represented by a unitary operator $\hat{C}_{4}$, which acts on $%
\mathcal{H}_{C}$ and that can be written as a $4\times 4$ matrix. Two
special cases that have been considered in the literature \cite%
{Mackay02,Tregenna03,Inui04(b)} are the Grover coin%
\begin{equation}
\hat{C}_{4,G}=\frac{1}{2}\left( 
\begin{array}{cccc}
-1 & 1 & 1 & 1 \\ 
1 & -1 & 1 & 1 \\ 
1 & 1 & -1 & 1 \\ 
1 & 1 & 1 & -1%
\end{array}%
\right) ,  \label{Grover}
\end{equation}%
and the DFT (discrete Fourier transform) coin%
\begin{equation}
\hat{C}_{4,DFT}=\frac{1}{2}\left( 
\begin{array}{cccc}
1 & 1 & 1 & 1 \\ 
1 & i & -1 & -i \\ 
1 & -1 & 1 & -1 \\ 
1 & -i & -1 & i%
\end{array}%
\right) .  \label{DFT}
\end{equation}

The state of the system after $n$ steps of the walk can be written as 
\begin{equation}
\left\vert \psi \right\rangle _{n}=\left( \hat{C}_{4}\hat{D}\right)
^{n}\left\vert \psi \right\rangle _{0},
\end{equation}%
with $\left\vert \psi \right\rangle _{0}$ the initial state of the system.
Finally, the probability distribution for the particle be at position $%
\left( x,y\right) $ after $n$ iterations is given by%
\begin{equation}
P\left( x,y;n\right) =\sum_{c\in \left\{ r,l,u,d\right\} }\left\vert
\left\langle x,y,c\right. \left\vert \psi \right\rangle _{n}\right\vert
^{2}=\sum_{c\in \left\{ r,l,u,d\right\} }P^{c}\left( x,y;n\right) ,
\label{probability}
\end{equation}%
with $P^{c}\left( x,y;n\right) =\left\vert c_{x,y}^{\left( n\right)
}\right\vert ^{2}$ the probability distributions for the particle be at
position $\left( x,y\right) $ and the coin in state $\left\vert
c\right\rangle $, $c\in \left\{ r,l,u,d\right\} $.

\section{Implementation}

In order to implement the two--dimensional QW one needs a walker that can
walk in two orthogonal directions, a plane, and a four--state qu$d$it. Here
we propose an implementation of this process that makes use of classical
resources only, following the same spirit as in \cite{Knight03,Knight03(b)}:
The four states of the coin will correspond to four different spatial paths
that the light field can follow (what, in the notation of \cite{Knight03(b)}%
, borrowed from \cite{Spreeuw01}, corresponds to a four--state position
cebit), and the walker role will be played by the field frequency, again as
in \cite{Knight03,Knight03(b)}, that can be increased or decreased in the
two orthogonal directions corresponding to two orthogonal polarization
states of the light field, say $\mathbf{x}$ and $\mathbf{y}$.

In Fig. 1 a schematic of the first step of the QW is schetched. In Fig. 1(a)
the four parallel light beams, which propagate along the z--axis and are
linearly polarized at $\pi /4$ with respect to the x--axis, first cross an
array of devices that perform the conditional displacement, Eqs. (\ref{D1}%
)--(\ref{D4}): The frequency of the $\mathbf{x}$--polarized ($\mathbf{y}$%
--polarized) light is increased or decreased in beams marked with $r$ or $l$
($u$ or $d$), respectively. Each of these devices can consist, e.g., of a
polarization beam--splitter (that separates the two--polarization components
of the incident beam, the frequency of one of which is suitably increased or
decreased by means of an electrooptic modulator), plus two mirrors and a
second polarization beam--splitter for recombining the two polarization
components back into a single beam after the frequency displacement. After
the implementation of $\hat{D}$, the four beams cross a second device in
which the $\hat{C}_{4}$ operation is implemented. Let us see how this
operation can be done.

In Fig. 2, a schematic of the device performing $\hat{C}_{4}$ is shown. The
four incoming beams suffer five transformations when crossing the $\hat{C}%
_{4}$ device. First, some phase is added to each of the fields, let us call
this operation $\hat{F}_{1}$, which is represented by the operator%
\begin{equation}
\hat{F}_{j}=\left( 
\begin{array}{cccc}
e^{i\phi _{j1}} & 0 & 0 & 0 \\ 
0 & e^{i\phi _{j2}} & 0 & 0 \\ 
0 & 0 & e^{i\phi _{j3}} & 0 \\ 
0 & 0 & 0 & e^{i\phi _{j4}}%
\end{array}%
\right) .  \label{Fi}
\end{equation}%
with $j=1$. After $\hat{F}_{1}$, beams $r$ and $l$ (and, separately, beams $%
u $ and $d$) are mixed in a beam splitter, let us call this operation $\hat{S%
}_{1}$, which in matrix form reads%
\begin{equation}
\hat{S}_{1}=\left( 
\begin{array}{cccc}
\cos \theta _{11} & i\sin \theta _{11} & 0 & 0 \\ 
i\sin \theta _{11} & \cos \theta _{11} & 0 & 0 \\ 
0 & 0 & \cos \theta _{12} & i\sin \theta _{12} \\ 
0 & 0 & i\sin \theta _{12} & \cos \theta _{12}%
\end{array}%
\right) .  \label{S1}
\end{equation}%
Then, the third step is similar to the first one, i.e., the phase of the
four beams are increased again. This is represented by the matrix Eq. (\ref%
{Fi}) with $j=2$. In the fourth step, similar to the second one, beams $r$
and $u$ (and, separately, beams $l$ and $d$) are mixed in a beam splitter,
let us call this operation $\hat{S}_{2}$. This is represented by 
\begin{equation}
\hat{S}_{2}=\left( 
\begin{array}{cccc}
\cos \theta _{21} & 0 & i\sin \theta _{21} & 0 \\ 
i\sin \theta _{21} & 0 & \cos \theta _{21} & 0 \\ 
0 & \cos \theta _{22} & 0 & i\sin \theta _{22} \\ 
0 & i\sin \theta _{22} & 0 & \cos \theta _{22}%
\end{array}%
\right) .  \label{S2}
\end{equation}%
The final step is a new dephasing of the beams, represented by Eq. (\ref{Fi}%
) with $j=3$. The global effect of these five operations is given by 
\begin{equation}
\hat{C}_{4}=\hat{F}_{3}\cdot \hat{S}_{2}\cdot \hat{F}_{2}\cdot \hat{S}%
_{1}\cdot \hat{F}_{1},
\end{equation}%
whose matrix elements can be writen as 
\begin{equation}
\hat{C}_{4}=\left( 
\begin{array}{cccc}
c_{11}c_{21}e^{i\alpha _{11}} & is_{11}c_{21}e^{i\alpha _{12}} & 
ic_{12}s_{21}e^{i\alpha _{13}} & -s_{12}s_{21}e^{i\alpha _{14}} \\ 
ic_{11}s_{21}e^{i\alpha _{21}} & -s_{11}s_{21}e^{i\alpha _{22}} & 
c_{12}c_{21}e^{i\alpha _{23}} & is_{12}c_{21}e^{i\alpha _{24}} \\ 
is_{11}c_{22}e^{i\alpha _{31}} & c_{11}c_{22}e^{i\alpha _{32}} & 
-s_{11}s_{22}e^{i\alpha _{33}} & ic_{12}s_{22}e^{i\alpha _{34}} \\ 
-s_{11}s_{22}e^{i\alpha _{41}} & ic_{11}s_{22}e^{i\alpha _{42}} & 
is_{12}c_{22}e^{i\alpha _{43}} & c_{12}c_{22}e^{i\alpha _{44}}%
\end{array}%
\right) ,  \label{Cs}
\end{equation}%
with $s_{ij}=\sin \theta _{ij}$ and $c_{ij}=\cos \theta _{ij}$. The phase
factors appearing in (\ref{Cs}) are related with the phase factors in (\ref%
{Fi}) through 
\begin{eqnarray}
\alpha _{11} &=&\phi _{11}+\phi _{21}+\phi _{31},\ \ \ \ \ \alpha _{12}=\phi
_{12}+\phi _{21}+\phi _{31}, \\
\alpha _{13} &=&\phi _{13}+\phi _{23}+\phi _{31},\ \ \ \ \ \alpha _{14}=\phi
_{14}+\phi _{23}+\phi _{31}, \\
\alpha _{21} &=&\phi _{11}+\phi _{21}+\phi _{32},\ \ \ \ \ \alpha _{22}=\phi
_{12}+\phi _{21}+\phi _{32}, \\
\alpha _{23} &=&\phi _{13}+\phi _{23}+\phi _{32},\ \ \ \ \ \alpha _{24}=\phi
_{14}+\phi _{23}+\phi _{32}, \\
\alpha _{31} &=&\phi _{11}+\phi _{22}+\phi _{33},\ \ \ \ \ \alpha _{32}=\phi
_{12}+\phi _{22}+\phi _{33}, \\
\alpha _{33} &=&\phi _{13}+\phi _{24}+\phi _{33},\ \ \ \ \ \alpha _{34}=\phi
_{14}+\phi _{24}+\phi _{33}, \\
\ \alpha _{41} &=&\phi _{11}+\phi _{22}+\phi _{34},\ \ \ \ \ \alpha
_{42}=\phi _{12}+\phi _{22}+\phi _{34}, \\
\alpha _{43} &=&\phi _{13}+\phi _{24}+\phi _{34},\ \ \ \ \ \alpha _{44}=\phi
_{14}+\phi _{24}+\phi _{34},\ \ \ \   \label{alfas}
\end{eqnarray}

Then, the operations performed for constructing $\hat{C}$ provide a class of
possible transformations, and depending on the values of parameters $\theta
_{ij}$ ($i,j=1,2$) and $\phi _{ij}$, through Eqs. (\ref{alfas}), different
transformations are obtained. For example, the Grover coin $\hat{C}_{4G}$,
Eq.(\ref{Grover}), is obtained by taking 
\begin{equation}
\theta _{11}=\theta _{12}=\theta _{21}=\theta _{22}=\pi /4,  \label{m}
\end{equation}%
for the beam splitters, and 
\begin{eqnarray}
\phi _{11} &=&\frac{\pi }{4},\ \ \ \   \notag \\
\phi _{12} &=&\phi _{14}=\phi _{31}=\phi _{34}=0,  \notag \\
\phi _{13} &=&-\phi _{21}=-\phi _{22}=\phi _{32}=\phi _{33}=\frac{\pi }{2},\
\   \notag \\
\ \phi _{23} &=&\phi _{24}=\pi ,
\end{eqnarray}%
for the phase filters. With respect to the DFT coin, Eq. (\ref{DFT}), it is
a little bit more complicated: By taking again (\ref{m}) for the beam
splitters and 
\begin{eqnarray}
\phi _{11} &=&\phi _{13}=\phi _{22}=\phi _{23}=\phi _{24}=0,  \notag \\
\phi _{12} &=&\phi _{14}=-\phi _{21}=\phi _{31}=\phi _{33}=-\frac{\pi }{2}, 
\notag \\
\phi _{32} &=&\phi _{34}=-\pi ,
\end{eqnarray}%
for the phase filters one obtains%
\begin{equation}
\hat{C}_{4,DFT}^{\prime }=\frac{1}{2}\left( 
\begin{array}{cccc}
1 & 1 & 1 & 1 \\ 
1 & 1 & -1 & -1 \\ 
1 & -1 & i & -i \\ 
1 & -1 & -i & i%
\end{array}%
\right) ,  \label{DFTbis}
\end{equation}%
which is very similar to Eq. (\ref{DFT}). In fact, the DFT matrix is
obtained from Eq. (\ref{DFTbis}) by making%
\begin{equation}
\hat{C}_{4,DFT}=\hat{A}\cdot \hat{C}_{4,DFT}^{\prime }\cdot \hat{A}^{-1},
\end{equation}%
with 
\begin{equation}
\hat{A}=\left( 
\begin{array}{cccc}
1 & 0 & 0 & 0 \\ 
0 & 0 & 1 & 0 \\ 
0 & 1 & 0 & 0 \\ 
0 & 0 & 0 & 1%
\end{array}%
\right) .  \label{A}
\end{equation}%
Notice that operator $\hat{A}$ interchanges indexes 2 and 3, what physicaly
means that the light beams $l$ and $u$ must be permuted at the entrance and
at the exit of the scheme in Fig. 1, what can be done by means of a Kepler
telescope.

Up to this point we have seen that a single step of the QW in two dimensions
can be performed by the device represented in Fig. 1. In order to perform $n$
steps, we only need to reinject the output of the device at its entrance.
This is readily achieved by using optical cavities (in Fig. 3 we show a
scheme of the complete setup). In the device, the initial condition is
chosen by fixing the phases and intensities of the four incident beams, and
at the cavity output, the frequency of the emerging field performs the
two--dimensional QW. Of course the output field spectrum must be analyzed,
with polarizers and frequency analizers, in order to extract the
two--dimensional QW: After passing a linear polarizer set to $0%
{{}^o}%
$ ($90%
{{}^o}%
$), from the spectrum of the polarized field one obtains $P\left(
x,0;n\right) $ ($P\left( 0,y;n\right) $), which suitably combined provide $%
P\left( x,y;n\right) $.

Let us note that the use of optical cavities imposses some restrictions (see 
\cite{Knight03(b)} for a more detailed discussion on these) as, e.g., the
intracavity field frequencies must resonate with the cavity modes, unless it
be a pulse with a duration shorter than the cavity roundtrip time. Also one
must take care that the optical paths of the different beams be equal and
that the polarization of the light field does not suffer variations along
the roundtrip (what prevents the use of optical fiber cavities). But these
technicalities can be readily solved.

Finally it is worth commenting that the device we are proposing here can
also implement the QW on the line with two coins, as recently proposed in
Ref. \cite{Inui04}. For that purpose, we only need to not distinguish
between the two polarization states of the light, i.e., the walk has to be
performed on a unique dimension, namely, the frequency of the field.

\section{Conclusion}

We have proposed an experimental setup for the implementation of the
two--dimensional QW. Our device consists of classical resources only and has
the advantage that the unitary transformation performed in it is tunable in
the sense that by modifying the parameters of the system, different unitary
transformations can be easily reproduced. The device we are proposing can be
generalized to implement the QW on the circle in either one or the two
dimensions by following the same technical solutions already proposed for
the one--dimensional QW \cite{Knight03(b)}.

The fact that the two--dimensional QW can be implemented by only classical
means suggests, as it was the case for the one--dimensional QW \cite%
{Knight03,Knight03(b)}, that it is a classical process in which nonlocal
entanglement plays no role. Recently \cite{Kendon05} this conclussion has
been discussed and we refer the reader to Ref. \cite{Kendon05} for more
details, as we are not going to discuss this here. Nevertheless, let us
emphasize that in higher dimensional QWs, e.g., the three--dimensional one,
quantum entanglement manifests in the amount of classical resources needed
for the implementation, as the implementation of the three necessary qubits
requires 8 light beams (in general, $n$ qubits would require $2^{n}$ light
beams \cite{Spreeuw01}). In this sense, the two-dimensional QW is the higher
dimensional one that can be implemented classically without a sensible
difference in the resources needed as compared with a \emph{quantum}
implementation.

This work has been financially supported by Spanish Ministerio de Ciencia y
Tecnolog\'{\i}a and European Union FEDER, Project BFM2002-04369-C04-01. We
gratefully acknowledge fruitful discussions with Germ\'{a}n J. de Valc\'{a}%
rcel.

\pagebreak

{\LARGE Figure Captions}

\textbf{Fig.1.} Schematic of the device performing a single step of the
two--dimensional QW. In the boxes marked with $+\omega _{x\left( y\right) }$
and $-\omega _{x\left( y\right) }$, the frequency of the $x$ $\left(
y\right) $ polarization component of the field in increased or decreased by
an amount $\omega _{x\left( y\right) }$. In the box marked with $\hat{C}_{4}$%
, a unitary transformation of the incoming vector $\left( r,l,u,d\right)
^{T} $ is performed (see Fig. 2 for details).

\textbf{Fig.2.} Schematic of the device performing the unitary
transformation $\hat{C}_{4}$. The boxes marked with $\phi _{ij}$ are
dephasing elements that increase the field phase in $\phi _{ij}$. The
rounded crossings indicate the presence of a beam--splitter.

\textbf{Fig.3.} Schematic of the optical cavity propopsed for implementing
the two--dimensional QW. The four optical paths are marked with a different
type of line for gguiding the eye. The black (grey) rectangles correspond to
perfectly (patially) reflecting mirrors.


\begin{thebibliography}{99}
\bibitem{Ambainis04} A. Ambainis, e-print quant-ph/0403120

\bibitem{Kempe03} J. Kempe, Contemp. Phys. \textbf{44}, 307 (2003)

\bibitem{Meyer96} D. Meyer, J. Stat. Phys. \textbf{85}, 551 (1996)

\bibitem{Aharonov93} Y. Aharonov, L. Davidovich, and N. Zagury, Phys. Rev. A 
\textbf{48}, 1687 (1993)

\bibitem{Fahri98} E. Fahri and S. Gutmann, Phys. Rev. A \textbf{58}, 915
(1998)

\bibitem{Nayak} A. Nayak, and A. Vishwanath, e-print quant-ph/0010117

\bibitem{Konno} N. Konno, Quantum Information Processing \textbf{1}, 345
(2002)

\bibitem{Carteret} H.A. Carteret, M.E.H. Ismail, and B. Richmond, J. Phys. A 
\textbf{36}, 8775 (2003)

\bibitem{Kendon03} V. Kendon and B. Tregenna, Phys. Rev. A \textbf{67},
042315 (2003)

\bibitem{Lopez03} C.C. L\'{o}pez and J.P. Paz, Phys. Rev. A \textbf{68},
052305 (2003)

\bibitem{Knight04} P.L. Knight, E. Rold\'{a}n, and J.E. Sipe, J. Mod. Opt. 
\textbf{51}, 1761 (2004)

\bibitem{Feldman04} E. Feldman and M. Hillery, Phys. Lett. A \textbf{324},
277 (2004)

\bibitem{Romanelli04} A. Romanelli, A.C. Sicardi Schifino, R. Siri, G. Abal,
A. Auyuanet, and R. Donangelo, Physica A \textbf{338}, 395 (2004)

\bibitem{Romanelli05} A. Romanelli, R. Siri, G. Abal, A. Auyuanet, and R.
Donangelo, Physica A \textbf{347}, 137 (2005)

\bibitem{Wojcik04} A. W\'{o}jcik, T. Lukzak, P. Kurzy\'{n}ski, A. Grudka,
and M. Bednarska, Phys. Rev. Lett. \textbf{93}, 180601 (2004)

\bibitem{Inui04} N. Inui and N. Konno, e-print quant-ph/0403153

\bibitem{Buerschaper04} O. Buerschaper and K. Burnett, e-print
quant-ph/0406039

\bibitem{Ribeiro04} P. Ribeiro, P. Milman, and R. Mosseri, e-print
quant-ph/0406071

\bibitem{Romanelli04(b)} A. Romanelli, A. Auyuanet, R. Siri, G. Abal, and R.
Donangelo, e-print quant-ph/0408183

\bibitem{Omar04} Y. Omar, N. Paunkovic, L. Sheridan, and S. Bose, e-print
quant-ph/0411065

\bibitem{Venegas04} S.E. Venegas--Andraca, J.L. Ball, K. Burnett, and S.
Bose, e-print quant-ph/0411151

\bibitem{Travaglione} B.C. Travaglione and G.J. Milburn, Phys. Rev. A 
\textbf{65}, 032310 (2002)

\bibitem{Dur} W. D\"{u}r, R. Raussendorf, V.M. Kendon, and H.-J. Briegel,
Phys. Rev. A \textbf{66}, 052319 (2002)

\bibitem{Sanders} B.C. Sanders, S.D. Bartlett, B. Tregenna, and P.L. Knight,
Phys. Rev. A \textbf{67}, 042305 (2003)

\bibitem{Zhao} Z. Zhao, J. Du, H. Li, T. Yang, Z.-B Chen, and J.-W. Pan,
e-print quant-ph/0212149

\bibitem{Di04} T. Di, M. Hillery, and M.S. Zubairy, Phys. Rev. A \textbf{70}%
, 032304 (2004)

\bibitem{Hillery} M. Hillery, J. Bergou, and E. Feldman, Phys. Rev. A 
\textbf{68}, 032314 (2003)

\bibitem{Knight03} P.L. Knight, E. Rold\'{a}n, and J.E. Sipe, Phys. Rev. A 
\textbf{68}, 020301(R) (2003)

\bibitem{Knight03(b)} P.L. Knight, E. Rold\'{a}n, and J.E. Sipe, Opt.
Commun. \textbf{227}, 147 (2003); erratum \textbf{232} (2004) 443

\bibitem{Jeong} H. Jeong, M. Paternostro, and M. S. Kim, Phys. Rev. A 
\textbf{69}, 012310 (2004)

\bibitem{Bouwmeester} D. Bouwmeester, I. Marzoli, G.P. Karman, W. Schleich,
and J.P. Woerdman, Phys. Rev. A \textbf{61}, 013410 (2000)

\bibitem{Kendon05} V. Kendon and B. Tregenna, Phys. Rev. A \textbf{71},
022307 (2005)

\bibitem{Mackay02} T.D. Mackay, S.D. Bartlett, S.L. Stephenson, and B.C.
Sanders, J. Phys. A \textbf{35}, 2745 (2002)

\bibitem{Tregenna03} B. Tregenna, W. Flanagan, R. Maile, and V. Kendon, New.
J. Phys. \textbf{5}, 83.1 (2003)

\bibitem{Inui04(b)} N. Inui, Y. Konishi, and N. Konno, Phys. Rev. A \textbf{%
69}, 052323 (2004)

\bibitem{Spreeuw01} R.J.C. Spreeuw, Phys. Rev. A \textbf{63}, 062302 (2001)
\end{thebibliography}
\end{document}